\documentclass[pss]{wiley2sp}
\usepackage{amsmath}

\usepackage{ifpdf}

\begin{document}

\title{Spectral flow for Aharonov--Bohm rings generated by zero-mass lines}

\titlerunning{Spectral flow}

\author{%
  Timur Tudorovskiy\textsuperscript{\Ast,\textsf{\bfseries 1}}
  Vladimir E. Nazaikinskii\textsuperscript{\Ast,\textsf{\bfseries 2},\textsf{\bfseries 3}},
  Mikhail I. Katsnelson\textsuperscript{\Ast,\textsf{\bfseries 1}}}

\authorrunning{Timur Tudorovskiy et al.}  

\mail{e-mail
  \textsf{t.tudorovskiy@science.ru.nl}, Phone:
  +31 (0)24 36 52712, Fax: +31 (0)24 36 52424}

\institute{%
  \textsuperscript{1}\,Institute for Molecules and Materials, Radboud University of Nijmegen, Heyendaalseweg 135, 6525AJ Nijmegen, The Netherlands\\
  \textsuperscript{2}\,A. Ishlinsky Institute for Problems in Mechanics, Moscow, Russia\\
  \textsuperscript{3}\,Moscow Institute of Physics and Technology, Dolgoprudnyi, Moscow District, Russia}

\abstract{
\abstcol{
We found that the spectrum of the linear dispersion mode localized
along the zero-mass line for variable mass Dirac fermions shows the
nonzero spectral flow as a function of the external magnetic field
$B$. This happens due to the Aharonov--Bohm effect, leading to the
linear dependence of every eigenvalue on $B$. The nonzero spectral
flow allows to use the magnetic field slowly varying in time to
control the energy of the linear dispersion mode.}{%
}}

\titlefigure[]{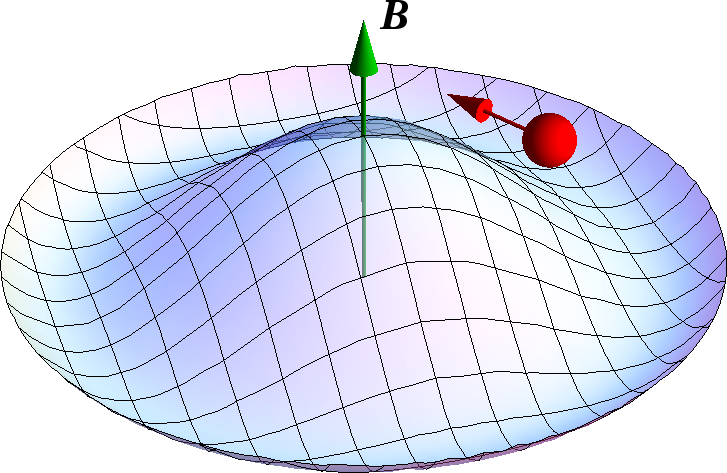}
\titlefigurecaption{
Illustration of the Aharonov--Bohm ring generated by the ZML. The particle is shown on the mass-square landscape}

\maketitle

\section{Introduction}
Massless Dirac fermions are charge carriers in graphene~\cite{kats12}
and topological insulators~\cite{qi11}, two very popular subjects in
contemporary physics. Geometric and topological concepts (see, e.g., Refs.~\cite{nak90,vol03}) are of crucial
importance for these systems. In particular, it was shown recently that nonzero spectral flow~\cite{APS3} is possible there~\cite{pro11,naz12}, which means a creation of electron-hole pairs from vacuum by adiabatically changing magnetic field~\cite{naz12}.
The works \cite{pro11,naz12} are rather formal. Here we give a simple physical example of a situation with nonzero spectral flow. It
is based on our previous consideration~\cite{tud12} of electron states
associated with zero-mass lines (ZML). These lines appear both in graphene
on hBN and other substrates and in topological insulators (e.g., in (Hg,Cd)Te/HgTe/(Hg,Cd)Te quantum wells)~\cite{tud12}.

The peculiarities of dynamics of two-dimensional Dirac fermions
localized in the vicinity of ZML were discussed
in many different aspects, see \cite{tud12} and references therein.
Classically it is described by the Hamiltonian function
\begin{equation}
H(\mathbf{p},\mathbf{r})=\sqrt{\mathbf{p}^2+m(\mathbf{r})^2},
\end{equation}
where $\mathbf{r}=\{x,\,y\}$ are the coordinates,
$\mathbf{p}=\{p_x,\,p_y\}$ are the corresponding momenta and $m$ is
the mass. The mass square landscape plays a role of the effective
potential if we choose $H(\mathbf{p},\mathbf{r})^2$ as an effective
Hamiltonian. Thus every ZML gives rise to a one-dimensional
channel, or waveguide, supporting the propagation of modes
localized in its vicinity. Every mode is characterized by the
transversal wave number related to the transversal energy. It turns
out that there is a mode corresponding to zero transversal energy.
This quasi-one-dimensional mode has a linear dispersion; i.e. its
energy $E$ of motion along the ZML is proportional to the
momentum.

Let us summarize the peculiarities of the linear dispersion mode
(LDM). Neglecting tunneling effects we can say that it is
unidirectional, meaning that along a given ZML only the motion in a
single direction is allowed. The Bohr--Sommerfeld quantization rule
giving the semiclassical answer for the spectral series associated
with the LDM reads
\begin{equation}
 \oint p_\tau d\tau = 2\pi n + \phi_B + \ldots,
\label{eq::BohrSom}
\end{equation}
where the integration should be performed along the ZML, $\tau$ is
the length of this line counted from some point, $p_\tau$ is the
longitudinal momentum, $n$ is an integer, $\phi_B=\pi$ for the
curve without self-intersections is the Berry phase locally
associated with the curvature and by $\ldots$ we denoted some terms
resulting from the displacement from the ZML (see \cite{tud12}). For the LDM we have
$p_\tau=E$, thus (\ref{eq::BohrSom}) takes the form $E
l=2\pi(n+1/2)+\ldots$, where $l$ is the length of the loop.

It is easy to modify the quantization condition for the case of non-zero external magnetic field. Indeed, the closed
ZML forms an Aharonov--Bohm ring in this case. Thus, an additional
magnetic flux $\Phi=B S$ through the loop bounding area $S$ should
be added on the right:
\begin{equation}
 E l = 2\pi (n+1/2) + \Phi + \ldots.
 \label{eq::AB}
\end{equation}
In this spectral series every eigenvalue linearly depends on the
magnetic flux. If $\Phi$ is changed (approximately) by a multiple
of $2\pi$, then the spectral series as a whole does not change at
all, except that the numbering of eigenvalues is shifted.
\begin{figure}
\begin{center}
\ifpdf
\includegraphics[width=4cm]{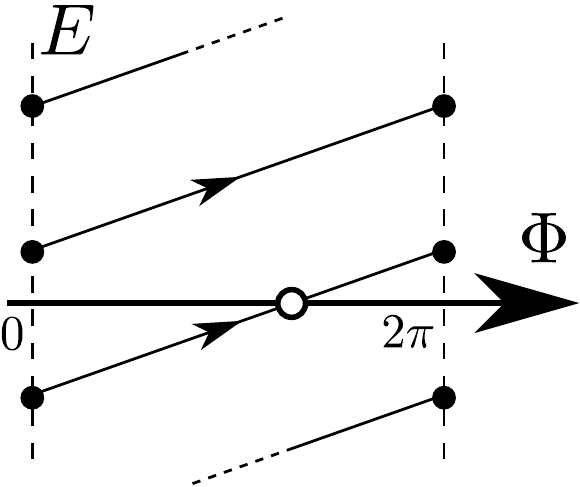}
\else
\includegraphics[width=4cm]{Figures/flow.eps}
\fi
 \caption{\label{fig::flow1}Illustration of the spectral flow associated
with the magnetic field. The spectral flow is defined as the number
of crossings of the horizontal line $E=0$ taken with the sign of
the derivative at the crossing point. In the figure spectral flow
is equal to one.}
\end{center}
\end{figure}
This shift in the numbering, which can equivalently be defined as
the number of crossings of the lines $E=E(\Phi)$ with the
horizontal axis $E=0$, is called the \textit{spectral flow}. Thus,
the LDM gives a nontrivial physical example of nonzero spectral
flow. Note that modes corresponding to nonzero
transversal energy do not contribute to the spectral flow, because
they always have a gap around zero. For the effective Hamiltonian
corresponding to the LDM on a closed ZML, switching on a magnetic
field whose magnetic flux is a multiple of $2\pi$ is equivalent to
a gauge transformation on the ZML. Of course, this transformation
cannot be extended to the entire domain bounded by the ZML if the
magnetic flux is nonzero. Thus, if we restrict ourselves to what
happens in the vicinity of the ZML, we can adopt the point of view
that the effective Hamiltonian does not change at all---we only
choose a different gauge. The above-mentioned gauge transformation
can be used to define a one-dimensional vector bundle on the
product of the ZML by the circle, and then the spectral flow
coincides with the index of the elliptic operator that is obtained
by adding the term $\partial/\partial t$ to the effective
Hamiltonian and acts on sections of this bundle, $t$ being the
coordinate on the circle.

If the magnetic field slowly varies in time, the energy levels will
still obey the dependence (\ref{eq::AB}). This permits controlling
the energy of the LDM by means of a magnetic field: an increase in
the magnetic field will lead to an increase in the energy and vice
versa.


Quantum-mechanically the dynamics of the Dirac fermion in the
magnetic field is described by the Hamiltonian
\begin{equation}
 H=\boldsymbol{\sigma}(\mathbf{p}-\mathbf{A})+\sigma_z m,
\end{equation}
where $\boldsymbol{\sigma}=\{\sigma_x,\,\sigma_y\}$ are Pauli
matrices and $\mathbf{A}$ is the vector potential. We consider a
uniform magnetic field oriented along $z$-axis
$\textbf{B}=\nabla\times\mathbf{A}$.

Let us consider a bent zero line $\gamma$ given by
$\{x,y\}=\mathbf{R}(\tau)$, where $\tau$ is a natural parameter,
i.e. $|\mathbf{R}'(\tau)|=1$. In the vicinity of this line we
introduce new variables $\tau$, $\xi$ by the equality
$\{x,y\}=\mathbf{R}(\tau)+\xi \mathbf{n}(\tau)$, where $\mathbf{n}$
is a unit normal vector on the curve at the point $\tau$. In the
curvilinear coordinates $\tau,\,\xi$ the Schr\"odinger equation
$H\Psi=E\Psi$ can be written as $\hat H\tilde \Psi=E\tilde \Psi$,
where $\tilde\Psi=\sqrt{1-k(\tau)\xi}\Psi$,
$k(\tau)=-(\mathbf{R}'(\tau)\mathbf{n}'(\tau))$ is the signed
curvature, and
\begin{eqnarray}
 \hat H&=&-\frac{i\boldsymbol{\sigma}\mathbf{R}'(\tau)}{\sqrt{1-\xi k(\tau)}}\frac{\partial}{\partial\tau}\frac{1}{\sqrt{1-\xi k(\tau)}}
 -i\boldsymbol{\sigma}\mathbf{n}(\tau)\frac{\partial}{\partial\xi}\nonumber\\
 &-&\frac{ik\boldsymbol{\sigma}\mathbf{n}(\tau)}{2(1-\xi k(\tau))}-\boldsymbol{\sigma}\mathbf{A}+\sigma_z m.
\label{eq::tilde-H}
\end{eqnarray}

The Hamiltonian (\ref{eq::tilde-H}) can be simplified in the
adiabatic approximation. We suppose that in the transverse
direction to $\gamma$ the wave function is localized at the scale
much smaller than the radius of curvature; i.e. $\xi k(\tau)\ll 1$.
On the other hand, it is natural to assume that the curvature does
not change significantly within this scale, which implies that $\xi
k'(\tau)/k(\tau)\ll 1$. We will consider the magnetic field $B$
corresponding to the magnetic length $l_0=1/\sqrt{B}$ comparable to
the radius of the curvature $1/k(\tau)$. This provides a magnetic
flux of the order of unity through the area bounded by the closed
curve $\gamma$.

In the vicinity of the ZML we have
$\mathbf{A}(\tau,\xi)=\mathbf{A}(\tau,0)+\mathbf{A}'_\xi(\tau,0)\xi+\ldots$
Introducing the notation $\mathbf{A}_0(\tau)=\mathbf{A}(\tau,0)$
and using the equality
$\mathbf{A}_0(\tau)=(\mathbf{A}_0\mathbf{R}')\mathbf{R}'(\tau)+(\mathbf{A}_0\mathbf{n})\mathbf{n}(\tau)$
we find $\hat H\simeq \hat H_0+\hat H_1$,
\begin{eqnarray}
\hat H_0&=&\boldsymbol{\sigma}\mathbf{R}'(\tau)\hat P_\tau
 +\boldsymbol{\sigma}\mathbf{n}(\tau)\hat P_\xi
 +\sigma_z m,\\
\hat H_1&=&\boldsymbol{\sigma}\mathbf{R}'(\tau)\xi k(\tau)\hat
p_\tau
 -\frac{ik}{2}\boldsymbol{\sigma}\mathbf{n}(\tau)-\boldsymbol{\sigma}\mathbf{A}'_\xi(\tau,0)\xi,
\label{eq::H0H1}
\end{eqnarray}
where $\hat P_\tau=\hat p_\tau-(\mathbf{A}_0\mathbf{R}')$, $\hat p_\tau=-i\partial/\partial\tau$, $\hat P_\xi=-i\partial/\partial\xi-(\mathbf{A}_0\mathbf{n})$. Let us choose a gauge
$\mathbf{A}=\{-By,0,0\}$. Then
$\boldsymbol{\sigma}\mathbf{A}'_\xi(\tau,0)=-\sigma_x B n_y(\tau)$.
Replacing the operator $\hat p_\tau$ in (\ref{eq::H0H1}) by the
variable $p_\tau$ we obtain symbols \cite{mas81} $H_0$ and $H_1$.


In the adiabatic approximation the effective dynamics along the line
$\gamma$ is described by the effective Hamiltonian $\hat L_B\simeq
\hat L_0+\hat L_1$. The symbols $L_0$ and $L_1$ can be found from
general considerations \cite{bel06}. As usual, $L_0$ is an
eigenvalue of $H_0$, i.e.
$H_0\chi=L_0\chi.$ 
The symbol $L_1$ is given by
\begin{eqnarray}
 L_1&=&\langle\chi^\dagger H_1\chi\rangle_\xi\nonumber\\
 &+&i\left<\chi^\dagger\frac{\partial L_0}{\partial\tau}\frac{\partial\chi}{\partial p_\tau}\right>_\xi
 -i\left<\chi^\dagger\frac{\partial H_0}{\partial p_\tau}\frac{\partial\chi}{\partial \tau}\right>_\xi,
\label{eq::L1gen}
\end{eqnarray}
where $\langle\cdot\rangle_\xi$ means the integration over $\xi$.
Introducing the notation $P_\tau=p_\tau-(\mathbf{A}_0\mathbf{R}')$
and the function $\chi_0$ by the equality
 $\chi=e^{i(\mathbf{A}_0\mathbf{n})\xi}\chi_0$
we reduce the calculation of $L_0$ to the case without magnetic
field~\cite{tud12}. It gives $L_0=-P_\tau$.
Similar to \cite{tud12}, we find from \eqref{eq::L1gen}
\begin{equation}
 L_1=G(\tau)  \left[k(\tau)P_\tau+B\right]
 +\frac{k(\tau)}{2},
\end{equation}
where $G(\tau)=2\langle\tilde\chi_{01}\xi\chi_{02}\rangle_\xi$ is the displacement
from ZML, $\tilde\chi_{01}=(n_x-in_y)\chi_{01}$ and $\tilde\chi_{01}$, $\chi_{02}$ are real. The explicit expression for $G(\tau)$ is given in Ref.~\cite{tud12}.
Thus, the effective longitudinal equation $\hat L_B\psi=E\psi$ reads
\begin{equation}
\left(-\hat P_\tau+G(\tau)
 \left[k(\tau)\hat P_\tau+B\right]
 +\frac{k}{2}\right)\psi=E\psi,
\label{eq::long}
\end{equation}
Equation (\ref{eq::long}) can be solved in the semiclassical approximation.
The solution is
\begin{multline}
 \psi=\exp\left(-iE\tau+i\int\mathbf{A}_0 d\mathbf{R}\right)\\
 \times\exp\left(i\int G(\tau)
 \left[B-k(\tau)E\right] d\tau
 +\frac{i\phi}2\right).
 \label{eq::psi}
\end{multline}
where $\phi_B=\phi/2$ is the Berry phase locally associated with the
curvature~\cite{tud12}, $d\phi=k(\tau)d\tau$. Here $\phi$ is
none other than the angle between the tangent to the curve
$\gamma$ and the $x$-axis, so that it gets an increment of $2\pi$
when going around $\gamma$. Using the answer (\ref{eq::psi}) it is
easy to find the quantization condition for the case in which
$\gamma$ is a closed curve:
\begin{multline}\label{20}
E\left(l+ \oint G(\tau)
 k(\tau) d\tau\right)\\=2\pi\left(n+\frac12\right)+B\left(S+ \oint G(\tau) d\tau\right),
\end{multline}
where the terms given by the circular integrals describe the small
corrections to the length $l$ of the contour $\gamma$ and the area
$S$ bounded by this contour owing to the displacement of the wave
function from $\gamma$. Thus, the increase of the total flux by flux
quantum corresponds to the replacement $n \to n+1$. As a result, the
spectral flow is equal to one, as illustrated in Fig. \ref{fig::flow1}.

Below we give a formal proof of this intuitively clear statement.
Assume that the magnetic flux
\begin{equation*}
    \Phi=B\left(S+ \oint G(\tau) d\tau\right)
\end{equation*}
is an integer multiple of $2\pi$, $\Phi=2\pi m$. Then it follows
from the quantization condition \eqref{20} that the spectrum of the
effective Hamiltonian $\widehat L_B$ is the same as for $B=0$. In
other words, $\widehat L_B$ and $\widehat L=\widehat L_B|_{B=0}$
are \textit{isospectral}. In fact, $\widehat L_B$ can be obtained
from $\widehat L$ by a gauge transformation,
    $\widehat L_B= U^{-1}\widehat L U,$
where
\begin{equation*}
    U=\exp \left(-i\int\mathbf{A}_0 d\mathbf{R}-B\int
    G(\tau)\,d\tau\right).
\end{equation*}
Consider the family of operators
\begin{equation*}
    \widehat L(t)=(1-t)\widehat L+ t\widehat L_B,\qquad 0\leq t\leq 1.
\end{equation*}
The spectral flow arising when the magnetic field $B$ is switched
on is just the spectral flow $\operatorname{sf} \widehat L(t)$ of
this family. The latter can be expressed as the index of some
elliptic operator. Namely, consider the first-order differential
operator $\partial/\partial t+\widehat L(t)$ on the product
$\gamma\times[0,1]$. We glue the endpoints of the interval $[0,1]$
together, so that the interval becomes the circle $S^1$ and this
product becomes the torus $\gamma\times S^1$. The operator
$\partial/\partial t+\widehat L(t)$ becomes an operator with
continuous coefficients on this product if we use the
transformation  $U$ to glue together the values of the functions on
$\gamma\times\{0\}$ ¨ $\gamma\times\{1\}$. One can readily verify
that this operator on $\gamma\times S^1$ is elliptic, and by
\cite{APS3}
\begin{equation*}
    \operatorname{sf} \widehat L(t)=\operatorname{ind}
    \left(\frac{\partial}{\partial t}+\widehat L(t)\right),
\end{equation*}
where `ind' denotes the index of the operator. Thus, the spectral
flow is equal to the number of zero-modes of the operator
$\partial/\partial t+\widehat L(t)$ minus that of the adjoint
operator, that is, one.


To conclude, we have shown that a closed zero-mass line naturally generates
the Aharonov--Bohm ring for the Dirac fermions if the external magnetic
field is applied. For the linear dispersion mode the Aharonov--Bohm
effect shifts the entire (sub)spectrum, leading to nonzero spectral
flow. This effect can be used to control the energy of the linear
dispersion mode by means of the magnetic field slowly changing in
time. The spectral flow can be computed in topological terms as the
index of an elliptic operator on the two-dimensional torus.

\bibliographystyle{pss}
\bibliography{thispaper}

\end{document}